\documentclass[12pt]{article}
\usepackage{amssymb}
\usepackage{cite}
\usepackage[]{hyperref}
\input xy
\xyoption{all}

\begin{document}

\begin{center}

{\large\bf (0,2) versions of exotic (2,2) GLSMs}

\vspace{0.2in}

Hadi Parsian, Eric Sharpe, Hao Zou

Dep't of Physics\\
Virginia Tech\\
850 West Campus Dr.\\
Blacksburg, VA  24061\\

{\tt varzi61@vt.edu},
{\tt ersharpe@vt.edu},
{\tt hzou@vt.edu}

$\,$

\end{center}

In this paper we extend work on exotic two-dimensional
(2,2) supersymmetric gauged linear sigma models (GLSMs) in which,
for example, geometries arise via nonperturbative effects,
to (0,2) theories, and in so doing find some novel
(0,2) GLSM phenomena.  For one example, we describe examples in which bundles
are constructed physically as cohomologies of short complexes involving
torsion sheaves, a novel effect not previously seen in (0,2) GLSMs.
We also describe examples related by RG flow in which the physical
realizations of the bundles are related by quasi-isomorphism, analogous
to the physical realization of quasi-isomorphisms in D-branes and derived
categories, but novel in (0,2) GLSMs.  Finally, we also discuss (0,2)
deformations in various duality frames of other examples.

\begin{flushleft}
February 2018
\end{flushleft}

\newpage

\tableofcontents

\newpage

\section{Introduction}

Over the last decade there have been numerous advances in understanding
two-dimensional (2,2) supersymmetric gauged linear sigma models (GLSMs).  
These have included nonperturbative
realizations of geometry in nonabelian (see {\it e.g.}
\cite{Hori:2006dk,Hori:2011pd})
and abelian (see {\it e.g.} \cite{Caldararu:2007tc,Caldararu:2017usq}) 
GLSMs, and
perturbative realizations of non-complete-intersections such as
Pfaffians \cite{Jockers:2012zr}, as well numerous advances in other
areas.
Two-dimensional (0,2) GLSMs have also seen a number of advances
over the last decade, but 
so far there has not been any work
applying nonperturbative geometric realizations
to (0,2) theories.

In this paper we begin to fill this gap, by describing some novel
properties of (0,2) GLSMs that result from considering nonperturbatively
realized geometries and other non-complete-intersections in (0,2)
rather than (2,2) settings.

For one example, we find examples of bundles constructed physically in
(0,2) GLSMs that involve short complexes of both bundles and skyscraper
sheaves, whereas previously all such physically-realized
monad constructions involved
short complexes of bundles only.
For another example, we find that bundles related by RG flow and dualities
are sometimes constructed physically by quasi-isomorphic complexes,
yielding a physical use for quasi-isomorphism outside of 
derived categories.
 
For another example, we find a physical realization of quasi-isomorphisms,
outside of physical realizations of derived categories
\cite{Sharpe:1999qz,Sharpe:2003dr},
relating monad constructions for (0,2) theories related by RG flow.

We begin in section~\ref{sect:br-double} by describing the physical
realization of tangent bundles of branched double cover constructions
first described in \cite{Hellerman:2006zs,Caldararu:2007tc}.  
These GLSM constructions are nonperturbative, in the sense that geometry
is not realized perturbatively as the critical locus of a superpotential.
These furnish the examples of bundles realized by extensions of torsion
sheaves.  We discuss both tangent bundles as well as (0,2) deformations of
the theories.  In broad brushstrokes, the rest of this paper concerns
(0,2) versions of the theories discussed in \cite{Caldararu:2017usq}.
In section~\ref{sect:veronese} we discuss physical realizations of the
tangent bundle of a Veronese embedding, and show how quasi-isomorphism
plays a role in relating presentations of tangent bundles in theories 
related by RG flow.  In section~\ref{sect:Segre} we discuss (0,2)
deformations of the Segre embeddings discussed in 
\cite{Caldararu:2017usq}, and again see that physical realizations of
tangent bundles in theories related by RG flow, are related mathematically
by quasi-isomorphisms.  In section~\ref{sect:grass-cap} we describe
(0,2) moduli of  
intersections $G(2,N) \cap G(2,N)$ in various duality frames, and for
completeness we conclude in section~\ref{sect:exs-anom-free} with a few
concrete examples of anomaly-free (0,2) models on
the Calabi-Yau $G(2,5) \cap G(2,5)$.

There are nonperturbatively-constructed geometries in both
nonabelian \cite{Hori:2006dk} as well as abelian \cite{Caldararu:2007tc}
GLSMs.
Unfortunately, we do not have a
simple realization of the tangent bundles for the nonabelian cases, and so we
do not discuss (0,2) deformations or tangent bundles in phases of
$G(2,N) \cap G(2,N)$ realized ala \cite{Hori:2006dk}.

Other work on two-dimensional (0,2) theories from just the past few months
includes
\cite{Gu:2017nye,Fiset:2017auc,Franco:2017lpa,
Closset:2017yte,Weigand:2017gwb,Bertolini:2017lcz,Couzens:2017nnr,
Dedushenko:2017osi,
Franco:2018qsc,Bertolini:2018now,Jardine:2018sft}.

\section{Tangent bundles of branched double covers}
\label{sect:br-double}

Ordinarily in (0,2) GLSMs \cite{Distler:1993mk,Distler:1995mi}, 
bundles are described as the cohomology
of a monad, a three-term complex of vector bundles on the ambient space,
in which each vector bundle corresponds to a set of massless worldsheet
fermions.

In this section we will discuss examples in which the tangent bundle
is realized physically in a different form, as an extension of a set of
skyscraper sheaves.  To our knowledge, the only previous cases in
which anything analogous was described were in
\cite{Distler:1996tj}; however, there the sheaves arose because the 
$E$ or $J$ maps failed to be injective or surjective, respectively,
whereas by contrast here one is getting torsion sheaves as part of the
original three-term complex.

We will analyze two examples from \cite{Caldararu:2007tc}.
This paper described examples of abelian GLSMs with exotic phases, in which
geometry was realized via nonperturbative effects, and geometries of different
phases were not birational to one another.  (See also \cite{Hori:2006dk} for
nonabelian examples with analogous properties.)
In broad brushstrokes, the examples in \cite{Caldararu:2007tc} describe,
in one phase, complete intersections of quadrics, and in another phase,
either branched double covers or noncommutative resolutions of
branched double covers.  We will restrict ourselves in this paper
to cases describing ordinary branched double covers and not noncommutative
resolutions.

\subsection{First example: branched covers of ${\mathbb P}^1$}

\subsubsection{(2,2) locus}

Our first example \cite{Caldararu:2007tc}[section 4.1]
is the GLSM for ${\mathbb P}^{2g+1}[2,2]$.
We first recall the (2,2) theory, and then will describe (0,2)
deformations.
This is a $U(1)$ gauge theory with matter
\begin{itemize}
\item $2g+2$ chiral superfields $\phi_i$ of charge $+1$,
\item $2$ chiral superfields $p_a$ of charge $-2$,
\end{itemize}
and superpotential
\begin{displaymath}
W \: = \:
\sum_a p_a G_a(\phi) \: = \:
\sum_{ij} \phi_i \phi_j A^{ij}(p),
\end{displaymath}
where $G_a(\phi)$ are a pair of quadric polynomials and
$A^{ij}(p)$ is a symmetric $(2g+2) \times (2g+2)$ matrix with entries
that are linear in the $p$s.

For large FI parameter $r \gg 0$, the analysis of this GLSM is standard,
and it describes a complete intersection of the two quadrics
$\{ G_a = 0\}$ in ${\mathbb P}^{2g+2}$.  Note that for $g=1$,
this is ${\mathbb P}^3[2,2]$, an elliptic curve.

For $r \ll 0$, the analysis of this example is more exotic.  The D terms
imply that not all the $p$'s can vanish, in which case the superpotential
acts as a mass matrix for the $\phi$ fields.  Naively, this phase then appears
to describe a sigma model on ${\mathbb P}^1$; however, since we know that for
$g=1$ the GLSM describes a Calabi-Yau, the $r \ll 0$ phase cannot describe
a non-Calabi-Yau, and so this cannot be the answer.  

To understand this
phase, we must take into account the fact that generically on the space of
$p$'s, the only massless fields have charge $2$ rather than one.
Theories with nonminimal charges -- equivalently, theories in which
nonperturbative sectors are restricted -- were analyzed in 
\cite{Pantev:2005zs,Pantev:2005rh,Pantev:2005wj}.  
In particular, \cite{Hellerman:2006zs} argued that in two-dimensional gauge
theories with nonminimal charges, the theory `decomposes' into a disjoint
union of theories.  In the present case, this means that generically 
on the space of $p$'s, the theory describes a double cover of 
${\mathbb P}^1$.  Further analysis \cite{Caldararu:2007tc} shows that this
is a branched double cover of ${\mathbb P}^1$, branched away from the
locus $\{ \det A = 0 \}$, which has degree $2g+2$.  This is precisely a
genus $g$ curve.  In particular, for $g=1$, both the $r \ll 0$ and
$r \gg 0$ phases describe an elliptic curve, exactly as expected.

As a consistency check, let us compare Witten indices.  To do this, we need
to take into account the discrete Coulomb vacua which exist in the $r \ll 0$
phase.  These arise as the solutions to
\begin{displaymath}
\sigma^{2g+2} (-2 \sigma)^{-2} (-2 \sigma)^{-2} \: = \: q,
\end{displaymath}
which has $2g-2$ solutions.  A genus $g$ Riemann surface has
$\chi = 2-2g$, so between the Higgs and Coulomb branches, we see that
altogether the Witten index of the $r \ll 0$ phase is
\begin{displaymath}
(2g-2) + (2-2g) \: = \: 0.
\end{displaymath}
It is straightforward to show that for the $r \gg 0$ phase, 
$\chi( {\mathbb P}^{2g+1}[2,2]) = 0$.  Hence both phases have the same
(vanishing) Witten index.  (See also \cite{Wong:2017cqs} for a more
detailed analysis of Witten indices in this and related examples.)

Now, let us turn to the physical realization of the tangent bundle
of the genus $g$ curve appearing in the $r \ll 0$ phase.
Locally over the space of $p$ vevs, the left-moving fermions include a 
left-moving gaugino $\lambda_-$, the superpartners $\psi_{p a}$ of the $p$
fields, and the superpartners $\psi_{\phi i}$ of the $\phi$ fields.
However, the latter are only massless at special points on the moduli space,
specifically the points where $\{ \det A = 0 \}$.
This suggests that the tangent bundle should be given as the cohomology
of the complex
\begin{equation}  \label{eq:22curve}
0 \: \longrightarrow \: {\cal O} \: \stackrel{E_a}{\longrightarrow} \:
{\cal O}(2)^2 \: \stackrel{*}{\longrightarrow} \: \oplus {\cal O}_p \:
\longrightarrow \: 0,
\end{equation}
where each ${\cal O}_p$ is a skyscraper sheaf.
The map $E_a \propto p_a$ arises from the usual analysis of (0,2) theories
\cite{Distler:1993mk,Distler:1995mi}.
We take the second map to be
\begin{displaymath}
* \: = \: \frac{\partial}{\partial p_a} \det A(p).
\end{displaymath}
This is determined by the need for this sequence to be a complex:
from homogeneity of the matrix $A$,
\begin{displaymath}
* \circ E_a \: = \: p_a \frac{\partial}{\partial p_a} \det A(p) \:
\propto \: \det A(p),
\end{displaymath}
which vanishes over the skyscraper sheaves above.

Mathematically, we can understand this\footnote{
We would like to thank T.~Pantev for explaining this to us.
} as a special case of the
Hurwitz formula, which can be described as follows.
Let $\pi: X \rightarrow S$ be a finite cover of smooth varieties,
and suppose that $\pi$ has simple ramification, meaning that the branch
divisor $B \subset S$ is smooth, and that over a neighborhood of each
point of $B$, the cover $\pi$ looks like a ramified cover plus
non-intersecting sheets.  Let $D \subset X$ denote the ramification divisor;
in the case of simple ramification, $\pi$ is an isomorphism between $D$ and $B$.
In this case, there is a short exact sequence
\begin{equation}   \label{eq:hurwitz}
0 \: \longrightarrow \: TX \: \longrightarrow \:
\pi^* TS \: \longrightarrow \:
i_* N_{D/X}  \: \longrightarrow \: 0,
\end{equation}
where $N_{D/X} = {\cal O}_X(D)|_D$ is the normal bundle of $D$ in $X$,
and $i: D \hookrightarrow X$ is inclusion.

In the present case, for $\Sigma$ the genus $g$ curve realized as a branched
double cover of ${\mathbb P}^1$, branched over the divisor $D$ consisting
of $2g+2$ points,
\begin{displaymath}
0 \: \longrightarrow \: T \Sigma \: \longrightarrow \:
\pi^* T {\mathbb P}^1 \: \longrightarrow \:
{\cal O}_D \: \longrightarrow \: 0.
\end{displaymath}
It is straightforward to compare this to our GLSM result above
(after normalizing the charges of the $p_a$ to be $1$ in mathematics
conventions, rather than $2$).
There, note that the cokernel of the map ${\cal O} \rightarrow
{\cal O}(2)^2$ is $\pi^* T {\mathbb P}^1$, so the we see that the GLSM 
sequence is equivalent to
\begin{displaymath}
\pi^* T {\mathbb P}^1 \: \longrightarrow \: {\cal O}_D \: 
\longrightarrow \: 0,
\end{displaymath}
which by virtue of the Hurwitz result above, has cohomology given by
$T \Sigma$, as desired.

In passing, we should mention there is an analogousr construction
of vector bundles described in \cite{donkron}[section 6.2.7],
as an extension of an ideal sheaf ${\cal I}$ rather than a torsion sheaf:
\begin{displaymath}
0 \: \longrightarrow \: {\cal O} \: \longrightarrow \:
{\cal E} \: \longrightarrow \: {\cal I} \: \longrightarrow \: 0.
\end{displaymath}
We do not claim to have a physical realization of this construction in
GLSMs, but thought it useful to mention the analogy.

\subsubsection{(0,2) deformations}

Before going on to our next example, let us pause to discuss
(0,2) deformations of the (2,2) theory above.
To review, so far we have discussed a (2,2) theory with superpotential
\begin{displaymath}
W \: = \:
\sum_a p_a G_a(\phi) \: = \: \sum_{ij} \phi_i \phi_j A^{ij}(p),
\end{displaymath}
for $G_a$ a set of quadrics in the $\phi$'s, and $A^{ij}$ a symmetric
$(2g+2) \times (2g+2)$ matrix, with entries linear in the $p$'s.
In (0,2) language, this would be described by
potential functions
\begin{displaymath}
E_i \: = \: - \sigma \phi_i, \: \: \:
E_a \: = \: 2 \sigma p_a,
\end{displaymath}
\begin{displaymath}
J_i \: = \: \sum_a p_a \frac{\partial G_a}{\partial \phi_i}, \: \: \:
J_a \: = \: G_a,
\end{displaymath}
where on the (2,2) locus, each $J$ is a derivative of $W$.

In principle, we can define a (0,2) deformation by replacing the $J_i$
above with
\begin{displaymath}
J_i \: = \: \sum_a p_a \left( \frac{\partial G_a}{\partial \phi_i} \: + \:
G_{ai}(\phi) \right),
\end{displaymath}
where the $G_{ai}$ are a set of (linear) functions of $\phi$ obeying
\begin{displaymath}
\sum_{a,i} \phi_i p_a G_{ai} \: = \: 0
\end{displaymath}
(so that $EJ = 0$ is obeyed).

Now, for convenience, define
\begin{displaymath}
B^{ij}(p) \: = \: \frac{1}{2} \sum_a p_a \frac{\partial}{\partial \phi_j}
G_{ai}(\phi),
\end{displaymath}
so that (since $G_{ai}$ is linear in $\phi$s)
\begin{displaymath}
J_i \: = \: 2 \sum_j \phi_j \left( A^{ij}(p) + B^{ij}(p) \right).
\end{displaymath}
Note that the potential term derived from $J_a$ is quartic in $\phi$s,
whereas the potential term derived from $J_i$ is quadratic in $\phi$s,
and so the quantity
$A^{ij}(p) + B^{ij}(p)$ acts as a mass matrix for the $\phi$s.

Now, let us consider the phases of this GLSM.
For $r \gg 0$, we have a (0,2) deformation of a complete intersection of 
quadrics, here ${\mathbb P}^{2g+1}[2,2]$.  The (0,2) deformation in this
phase acts as a modification of the left-moving gauge bundle.

For $r \ll 0$, the analysis is also very similar to the (2,2) locus,
except that because the mass matrix is $A+B$ instead of just $A$,
the branched double cover of the space of $p$'s is branched over the locus
\begin{displaymath}
\det( A + B) = 0 ,
\end{displaymath}
rather than the locus $\{ \det A=0\}$.  Thus we see that the (0,2) deformation
of the complete intersection has, as its $r \ll 0$ phase, a slightly
different geometry than one would have obtained on the (2,2) locus.

Such a result is not unusual in (0,2) theories, where the phases are
determined by the gauge bundle rather than the complete intersection
{\it per se} \cite{Distler:1993mk,Distler:1995mi}.

\subsection{Second example: branched covers of ${\mathbb P}^2$}

Our second example, from \cite{Caldararu:2007tc}[section 2.8], involves 
the GLSM for ${\mathbb P}^5[2,2,2]$.  This is a $U(1)$ gauge theory with
\begin{itemize}
\item 6 chiral superfields $\phi_i$ of charge $1$,
\item 3 chiral superfields $p_a$ of charge $-2$,
\end{itemize}
and a superpotential
\begin{displaymath}
W \: = \: \sum_a p_a G_a(\phi) \: = \:
\sum_{ij} \phi_i \phi_j A^{ij}(p),
\end{displaymath}
where the $G_a$ are quadric polynomials and $A^{ij}$ is a symmetric
$6 \times 6$ matrix with entries linear in the $p$'s.

For large FI parameter $r \gg 0$, the analysis is standard and the GLSM
describes the complete intersection ${\mathbb P}^5[2,2,2]$,
which is a K3 surface.

The analysis of the other phase, $r \ll 0$, proceeds as above.
From the D terms, the $p$'s are not all zero, hence the superpotential
defines a mass matrix for the $\phi_i$ over the space of $p$'s, a 
${\mathbb P}^2$.  Because at generic points the $p$'s are nonminimally
charged and the only massless fields, physics sees a branched double cover
of ${\mathbb P}^2$, branched over the degree six locus $\{ \det A = 0 \}$.
Such a branched double cover is another K3 surface, and so we see that
both phases in this model correspond to K3 surfaces.

Proceeding as before, the left-moving fermions describe the tangent
bundle as the cohomology of the short complex
\begin{equation}  \label{eq:22k3}
0 \: \longrightarrow \: {\cal O} \: \stackrel{E_a}{\longrightarrow} \:
{\cal O}(2)^3 \: \stackrel{*}{\longrightarrow} \:
{\cal O}(12) \otimes {\cal O}_D \: \longrightarrow \: 0,
\end{equation}
where 
\begin{displaymath}
E_a \: = \: p_a, \: \: \:
* \: = \: \frac{\partial}{\partial p_a} \det A(p),
\end{displaymath}
The left-most ${\cal O}$ corresponds to the left-moving gaugino,
the middle ${\cal O}(2)^3$ from the superpartners of the $p_a$,
and the right-most term from the superpartners of the $\phi_i$,
massless only along the locus $D \equiv \{ \det A = 0 \}$.
This is a complex due to homogeneity of the matrix $A^{ij}(p)$:
\begin{displaymath}
* \circ E_a \: = \: p_a \frac{\partial}{\partial p_a} \det A(p) \: \propto
\: \det A(p),
\end{displaymath}
which vanishes along $D$.
As before, the superpartners of the $\phi_i$ are not themselves charge $12$
objects, but correspond to a term coupling to the line bundle ${\cal O}(12)$
ultimately because they are only supported along the locus $D$.

Now, let us compare to the mathematics prediction.
In this case, the Hurwitz formula~(\ref{eq:hurwitz}) says
\begin{displaymath}
0 \: \longrightarrow \: T({\rm K3}) \: \longrightarrow \:
\pi^* T {\mathbb P}^2 \: \longrightarrow \:
\left( \pi^* {\cal O}(6) \right)
\otimes {\cal O}_D \: \longrightarrow \: 0.
\end{displaymath}
Normalizing the charge of $p_a$ to be $1$ instead of $2$,
we see that 
the sequence~(\ref{eq:22k3}) above matches the Hurwitz prediction for this
case.

\section{Quasi-isomorphism and the tangent bundle of Veronese embeddings}
\label{sect:veronese}

In this section we will see examples of theories related by RG flow
in which the physical realizations of the tangent bundles are related
mathematically by quasi-isomorphisms, a trick previously only seen in
discussions of D-branes and derived categories.

Consider a Veronese embedding of degree $d$, mapping ${\mathbb P}^n$ to
a projective space of dimension
\begin{displaymath}
N \: = \: \left( \begin{array}{c} n+d \\ d \end{array} \right) - 1.
\end{displaymath}
The corresponding GLSM \cite{Caldararu:2017usq} is a $U(1)$ gauge theory
with matter:
\begin{itemize}
\item $n+1$ chiral superfields $x_i$ of charge 1,
\item $N+1$ chiral superfields $y_{i_1 \cdots i_d}$ (symmetric in their
indices) of charge $d$,
\item $N+1$ chiral superfields $p_{i_1 \cdots i_d}$ (symmetric in their
indices) of charge $d$,
\end{itemize}
with superpotential
\begin{displaymath}
W \: = \: p_{i_1 \cdots i_d} \left( y_{i_1 \cdots i_d} - 
x_{i_1} \cdots x_{i_d} \right).
\end{displaymath}

Now, the geometry described by this GLSM is technically the
graph of the Veronese embedding, which is isomorphic to the original
${\mathbb P}^n$.  This projective space by itself does not have any
tangent bundle deformations; however, the physical realization of its
tangent bundle is related to that of ${\mathbb P}^n$ by quasi-isomorphism,
a relationship ordinarily only encountered in derived categories
\cite{Sharpe:1999qz,Sharpe:2003dr}.

Specifically, the tangent bundle is realized in the GLSM above as
the cohomology of the following monad over ${\mathbb P}^n$:
\begin{equation}  \label{eq:1st-pn-presentation}
0 \: \longrightarrow \: \stackrel{E}{\longrightarrow} \: 
{\cal O}(1)^{n+1} \oplus {\cal O}(d)^{N+1} \: \stackrel{J}{\longrightarrow}
\: {\cal O}(d)^{N+1} \: \longrightarrow \: 0,
\end{equation}
where the left-most ${\cal O}$ corresponds to the gaugino $\lambda_-$,
the middle bundle corresponds to the superpartners of $x_i$, $y_{i_1 \cdots
i_d}$, and the right-most bundle ${\cal O}(d)^{N+1}$ corresponds to the
superpartners of the $p_{i_1 \cdots i_d}$, and
\begin{displaymath}
E \: = \: (x_i, y_{i_1 \cdots, i_d}), \: \: \:
J \: = \: ( - x_{i_1} \cdots x_{i_{d-1}}, 1).
\end{displaymath}
($EJ = 0$ along the critical locus of the superpotential, namely the
${\mathbb P}^n.$)
Because of the presence of the identity maps in the $J$'s, arising from
\begin{displaymath}
J_{y_{j_1 \cdots j_d}} \: = \: \frac{\partial}{\partial y_{j_1 \cdots j_d} }
\left( y_{i_1 \cdots i_d} - x_{i_1} \cdots x_{i_d} \right),
\end{displaymath}
the same tangent bundle is obtained from the cohomology of the complex
\begin{equation}  \label{eq:2nd-pn-presentation}
0 \: \longrightarrow \: {\cal O} \: \stackrel{x_i}{\longrightarrow} \:
{\cal O}(1)^{n+1}.
\end{equation}

Mathematically, the two complexes~(\ref{eq:1st-pn-presentation})
and (\ref{eq:2nd-pn-presentation}) are said to be quasi-isomorphic,
as claimed.  In the next section we will see further examples of
quasi-isomorphisms relating physical realizations of bundles in theories
related by RG flow.

\section{Deformations of tangent bundles of Segre embeddings}
\label{sect:Segre}

The Segre embedding is an embedding of a product
${\mathbb P}^n \times {\mathbb P}^m$ in a higher-dimensional projective
space.  Mathematically, it is the map
\begin{displaymath}
        s: \ \mathbb{P}^n \times \mathbb{P}^m \rightarrow \mathbb{P}^{(n+1)(m+1)-1}
\end{displaymath}
defined by
\begin{displaymath}
        [x_0,\dots,x_n]\times [y_0,\dots,y_m] \mapsto [x_0 y_0, x_0y_1,\dots,x_n y_m ].
\end{displaymath}

A (2,2) GLSM realizing the Segre embedding was described in 
\cite{Caldararu:2017usq}, and is given as follows.
It is a $U(1) \times U(1)$ gauge theory with matter
\begin{itemize}
        \item $n+1$ chiral superfields $x_i$ of charge $(1,0)$,
        \item $m+1$ chiral superfields $y_i$ of charge $(0,1)$,
        \item $(n+1)(m+1)$ chiral superfields $z_{ij}$ of charge $(1,1)$,
        \item $(n+1)(m+1)$ chiral superfields $p_{ij}$ of charge $(-1,-1)$,
\end{itemize}
and with superpotential
\begin{displaymath}
        W = \sum_{i,j}p_{ij}G_{ij}(x,y,z) = \sum_{i,j}p_{ij}(z_{ij} - x_i y_j).
\end{displaymath}
We will see momentarily that physics realizes the RG flow from this
model to that for ${\mathbb P}^n \times {\mathbb P}^m$, and its
(0,2) deformations, via a quasi-isomorphism, just as in the last
section.

Now, the tangent bundle of ${\mathbb P}^n \times
{\mathbb P}^m$ admits deformations, and this is in fact used as
a canonical example in discussions of quantum sheaf cohomology
in (0,2) theories, see for example \cite{Katz:2004nn,McOrist:2007kp,McOrist:2008ji,Donagi:2011uz,Donagi:2011va,Closset:2015ohf}.
Mathematically, these deformations of the tangent bundle are given as
a cokernel ${\cal E}$, where
\begin{displaymath}
0 \: \longrightarrow \: {\cal O}^2 \: \stackrel{*}{\longrightarrow} \:
{\cal O}(1,0)^{n+1} \oplus {\cal O}(0,1)^{m+1} \: \longrightarrow \:
{\cal E} \: \longrightarrow \: 0,
\end{displaymath}
with
\begin{displaymath}
* \: = \: \left[ \begin{array}{cc}
A x & B x \\
C y & D y \end{array} \right],
\end{displaymath}
with $A, B$ $(n+1) \times (n+1)$ matrices and
$C, D$ $(m+1) \times (m+1)$ matrices.  In effect, this is a deformation of
two copies of the Euler sequences for the tangent bundles of the
two separate projective spaces, reducing to the tangent bundle in the
special case that $A$, $D$ are the identity and $B = 0 = C$.
Physically, in a (0,2) GLSM, the map $*$ is realized in the $E$ potentials
associated to the Fermi superfields associated with 
${\cal O}(1,0)^{n+1} \oplus {\cal O}(0,1)^{m+1}$.

The (0,2) deformations above also exist in the Segre embedding,
as expected.  For completeness, we list them here:
Define
\begin{displaymath}
E_{xi} = \sigma_1 (A x)_i + \sigma_2 (B x)_i, \: \: \:
E_{yj} = \sigma_1 (C y)_j + \sigma_2 (D y)_j,
\end{displaymath}
\begin{displaymath}
J_{xi} = - p^{ij} y_j, \: \: \:
J_{yj} = - p^{ij} x_i,
\end{displaymath}
\begin{displaymath}
E_{zij} = \sigma_1 ( A_{ik} z_{kj} + C_{jk} z_{ik} ) + \sigma_2(
B_{ik} z_{kj} + D_{jk} z_{ik} ), \: \: \:
J_{zij} = p^{ij},
\end{displaymath}
\begin{displaymath}
E_{pij} = - \sigma_1( A_{ki} p^{kj} + C_{kj} p^{ik}) - \sigma_2(
B_{ki} p^{kj} + D_{kj} p^{ik} ), \: \: \:
J_{pij} = z_{ij} - x_i y_j.
\end{displaymath}
One can show that $E J = 0$.

Next, let us compare complexes.  Recall
the analogue of the Euler complex for the deformation ${\cal E}$ of
the tangent bundle of ${\mathbb P}^n \times {\mathbb P}^m$ has the form
\begin{displaymath}
0 \: \longrightarrow \: {\cal O}^2 \: \stackrel{E}{\longrightarrow} \:
{\cal O}(1,0)^{n+1} \oplus
{\cal O}(0,1)^{m+1} \: \longrightarrow \:
{\cal E} \: \longrightarrow \: 0.
\end{displaymath}
The analogous complex for the tangent bundle deformation of the
Segre embedding is
\begin{displaymath}
0 \: \longrightarrow \: {\cal O}^2 \: \stackrel{E'}{\longrightarrow} \:
{\cal O}(1,0)^{n+1} \oplus
{\cal O}(0,1)^{m+1} \oplus
{\cal O}(1,1)^{(n+1)(m+1)} \: \stackrel{J}{\longrightarrow} \:
{\cal O}(1,1)^{(n+1)(m+1)} \: \longrightarrow \: 0.
\end{displaymath}
In this case, the tangent bundle deformation ${\cal E}$ is the
cohomology of this complex.  It is straightforward to check that the
complex above is quasi-isomorphic to the complex
\begin{displaymath}
0 \: \longrightarrow \: {\cal O}^2 \: \stackrel{E}{\longrightarrow} \:
{\cal O}(1,0)^{n+1} \oplus
{\cal O}(0,1)^{m+1} ,
\end{displaymath}
and so we see again that quasi-isomorphism is the mathematical
realization of RG flow in (0,2) theories, just as it is in
the physical realization of derived categories.

\section{(0,2) deformations of $G(2,N) \cap G(2,N)$}
\label{sect:grass-cap}

In \cite{Caldararu:2017usq}, a (2,2)
GLSM was given for the Calabi-Yau constructed as
the self-intersection of the Grassmannian $G(2,5)$, as well as several
dual descriptions of that GLSM.  In this section, we will describe the
deformations of that GLSM in its various duality frames, and compare
the results.

\subsection{First description}

The first (2,2) GLSM for $G(2,N) \cap G(2,N)$, presented in
\cite{Caldararu:2017usq}[section 4.1], was as a 
\begin{displaymath}
\frac{
U(1) \times SU(2) \times SU(2)
}{
{\mathbb Z}_2 \times {\mathbb Z}_2
}
\end{displaymath}
gauge theory with matter
\begin{itemize}
\item $N$ chiral multiplets $\phi^i_a$ in the $({\bf 2},{\bf 1})_1$
representation,
\item $N$ chiral multiplets $\tilde{\phi}^j_{a'}$ in the $({\bf 1},{\bf 2})_1$
representation,
\item $(1/2)N(N-1)$ chiral multiplets $p_{i j} = - p_{j i}$ in the
$({\bf 1},{\bf 1})_{-2}$ representation,
\end{itemize}
with superpotential
\begin{displaymath}
W \: = \: \sum_{i<j} p_{ij} \left( f^{ij}(B) - \tilde{B}^{ij} \right),
\end{displaymath}
where 
\begin{displaymath}
B^{ij} = \epsilon^{ab} \phi^i_a \phi^j_b, \: \: \:
\tilde{B}^{ij} = \epsilon^{a'b'} \tilde{\phi}^i_{a'} \tilde{\phi}^j_{b'}
\end{displaymath}
are the baryons in each $SU(2)$ factor, and $f^{ij}(x)$ define a 
linear isomorphism on the homogeneous coordinates of
${\mathbb P}^{(1/2)N(N-1)-1}$, defining the deformation of one of the
copies of the Pl\"ucker embedding.  Put another way,
\begin{displaymath}
f^{i j}(B) \: = \: f^{ij}_{k\ell} B^{k \ell}
\end{displaymath}
for a constant invertible matrix $f^{ij}_{k\ell}$.
Each ${\mathbb Z}_2$ factor in the gauge group linked the center of
one of the two $SU(2)$'s with a ${\mathbb Z}_2$ subgroup of $U(1)$,
and it is straightforward to check that the matter is invariant.

In this section, we shall describe (0,2) deformations of this theory.

The tangent bundle defined implicitly by this GLSM in its $r \gg 0$ phase
is given\footnote{
See {\it e.g.} \cite{Jia:2014ffa} for a discussion of physical realizations of
tangent bundles of PAX and PAXY
models, which form the prototype for this observation.
} by the
cohomology of the sequence
\begin{displaymath}
0 \: \longrightarrow \:
{\cal O} \oplus {\cal O}({\bf 3},{\bf 1})_0 \oplus {\cal O}({\bf 1},{\bf 3})_0
\: \stackrel{E}{\longrightarrow} \:
{\cal O}({\bf 2},{\bf 1})_1^{\oplus N} \oplus
{\cal O}({\bf 1},{\bf 2})_1^{\oplus N} \:
\stackrel{J}{\longrightarrow} \:
{\cal O}({\bf 1},{\bf 1})_2^{\oplus (1/2)N(N-1)} \: \longrightarrow \: 0.
\end{displaymath}
In our notation, ${\cal O}({\bf n},{\bf m})_p$ is the bundle defined
by representation ${\bf n}$ of the first $SU(2)$, ${\bf m}$ of the
second $SU(2)$, and charge $p$ of the $U(1)$ factor.
The leftmost factor,
${\cal O}\oplus {\cal O}({\bf 3},{\bf 1})_0 \oplus
{\cal O}({\bf 1},{\bf 3})_0$, is defined by the gauginos in the theory.
The middle factor is defined by the chiral multiplets
$\phi^i_a$, $\tilde{\phi}^i_{a'}$.  The rightmost factor is defined by
the $p_{ij}$.  As a consistency check, 
note that the rank of the resulting bundle is given by
\begin{displaymath}
2N + 2 N - 7 - (1/2)N(N-1)
\end{displaymath}
coinciding with the expected dimension given in
\cite{Caldararu:2017usq}[section 3.2.1].

The $r \ll 0$ phase is realized nonperturbatively in the form
of \cite{Hori:2006dk}, so as mentioned in the introduction, we shall not try
to write down a purely mathematical description of the tangent bundle.

Next, we consider $(0,2)$ deformations.  The $E$-terms are
\begin{eqnarray*}
        E_{p_{i_1 i_2}} & =& -\sigma \left( \tilde{N}^j_{i_1} \, p_{j i_2}
- \tilde{N}^j_{i_2} \, p_{j i_1} \right) , \\
        E_{a}^i & =&  \sigma^{b}_{a} \phi_{b}^i
+ N^i_j \sigma \phi_{a}^j, 
\\
\tilde{E}_{a'}^i & = & \tilde{\sigma}^{b'}_{a'} \tilde{\phi}_{b'}^i 
+ \tilde{N}^i_j \sigma \tilde{\phi}_{a'}^j,
\end{eqnarray*}
where $N^i_j$, $\tilde{N}^i_j$ are related by the constraint
\begin{equation}  \label{eq:first:constr}
N^k_{[j_1} f^{i_1 i_2}_{j_2] k} \: = \:
\tilde{N}^{[i_1}_k f^{i_2] k}_{j_1 j_2},
\end{equation}
with $J$ terms
\begin{eqnarray*}
        J_{p_{i_1 i_2}} & =& f^{i_1 i_2}_{j_1 j_2} B^{j_1 j_2} - \tilde{B}^{i_1 i_2},\\
        J_{\phi_{a}^k} & =& p_{i_1 i_2} f^{i_1 i_2}_{j_1 j_2} \frac{\partial B^{j_1 j_2}}{\partial \phi_{a}^{k}}, \\
 J_{\tilde{\phi}_{a'}^k} & =& 
 - p_{i_1 i_2} \frac{\partial \tilde{B}^{i_1 i_2}}{\partial \tilde{\phi}_{a'}^{k}},
\end{eqnarray*}
where $\sigma^{a}_{b}$ is traceless.
It is straightforward to check that $E J=0$, as required by supersymmetry.

The (2,2) locus is given by taking 
\begin{displaymath}
N^i_j \: = \: \delta^i_j \: = \: \tilde{N}^i_j,
\end{displaymath}
which is easily checked to satisfy condition~(\ref{eq:first:constr}).

As a demonstration that other solutions to constraint~(\ref{eq:first:constr})
exist, the reader can verify that in the case
\begin{displaymath}
f^{i_1 i_2}_{j_1 j_2} \: = \: \frac{1}{2} \left(
\delta^{i_1}_{j_1} \delta^{i_2}_{j_2} \: - \:
\delta^{i_1}_{j_2} \delta^{i_2}_{j_1} \right),
\end{displaymath}
constraint~(\ref{eq:first:constr}) is satisfied for
\begin{displaymath}
N^i_j \: = \: \alpha \delta^i_1 \delta^2_j \: = \:
\tilde{N}^i_j,
\end{displaymath}
where $\alpha$ is a constant.

\subsection{Double dual description}

Next, we turn to the `double dual' of this GLSM described in
\cite{Caldararu:2017usq}[section 4.2], obtained by dualizing both of the
$SU(2)$ factors in the GLSM for $G(2,N) \cap G(2,N)$ using the duality
described in \cite{Hori:2011pd}.  The result is a
\begin{displaymath}
\frac{
U(1) \times Sp(N-3) \times Sp(N-3)
}{
{\mathbb Z}_2 \times {\mathbb Z}_2
}
\end{displaymath}
gauge theory with
\begin{itemize}
\item $N$ fields $\varphi_i^a$ in the $({\bf N-3},{\bf 1})_{-1}$ representation,
\item $N$ fields $\tilde{\varphi}^{a'}_i$ in the $({\bf 1},{\bf N-3})_{-1}$
representation,
\item $(1/2)N(N-1)$ fields $b^{ij} = - b^{ji}$ in the
$({\bf 1},{\bf 1})_{2}$ representation,
\item $(1/2)N(N-1)$ fields $\tilde{b}^{ij} = - \tilde{b}^{ji}$ in the
$({\bf 1},{\bf 1})_{2}$ representation,
\item $(1/2)N(N-1)$ fields $p_{ij} = - p_{ji}$ in the
$({\bf 1},{\bf 1})_{-2}$ representation,
\end{itemize}
with superpotential
\begin{eqnarray*}
W & = & \sum_{i<j} p_{ij} \left( f^{ij}(b) - \tilde{b}^{ij}\right) \: + \:
\varphi^a_i \varphi^b_j J_{ab} b^{ij} \: + \:
\tilde{\varphi}^{a'}_i  \tilde{\varphi}^{b'}_j J_{a'b'} \tilde{b}^{ij}, \\
& = & \left( A(p)_{ij} + \varphi^a_i \varphi^b_j J_{ab} \right) b^{ij} \: + \:
\left( C(p)_{i j} + \tilde{\varphi}^{a'}_i \tilde{\varphi}^{b'}_j 
J_{a' b'} \right) \tilde{b}^{ij},
\end{eqnarray*}
where $J$ is the antisymmetric symplectic form, and $A(p)$, $C(p)$ are
matrices that can be derived from the first line of the expression for the
superpotential.

The $r \gg 0$ phase realizes geometry nonperturbatively in the sense of
\cite{Hori:2006dk}, so as described in the introduction, we shall not
try to write down a purely mathematical description of the tangent bundle.
The $r \ll 0$ phase, on the other hand, can be described perturbatively.

In the $r \ll 0$ phase, D terms imply that not all of the
$\varphi^a_i$, $\tilde{\varphi}^{a'}_i$, and $p_{ij}$ can vanish.  
The tangent bundle is built physically\footnote{
See {\it e.g.} \cite{Jia:2014ffa} for a discussion of physical realizations of
tangent bundles of PAX and PAXY
models, which form the prototype for this observation.
} as the cohomology of the complex
\begin{displaymath}
0 \: \longrightarrow \:
{\cal O} \oplus {\cal O}(({\rm adj},{\bf 1})_0) \oplus
{\cal O}({\bf 1},{\rm adj})_0) \:
\stackrel{E}{\longrightarrow} \:
A
\: \stackrel{J}{\longrightarrow} \:
{\cal O}( ({\bf 1},{\bf 1})_{-2})^{\oplus (2) (1/2) N (N-1) } 
\: \longrightarrow \: 0,
\end{displaymath}
where
\begin{displaymath}
A \: = \:
{\cal O}( ({\bf N-3},{\bf 1})_{-1})^{\oplus N} \oplus
{\cal O}(({\bf 1},{\bf N-3})_{-1})^{\oplus N} \oplus
{\cal O}(({\bf 1},{\bf 1})_{-2})^{\oplus (1/2)N(N-1)} .
\end{displaymath}
The leftmost terms are from the gauginos, the middle terms ($A$) are
from $\varphi^a_i$, $\tilde{\varphi}^{a'}_i$, and $p_{ij}$,
and the rightmost terms are from $b^{ij}$,
$\tilde{b}^{ij}$.
As a consistency check, note that for $N=5$ the rank of this bundle is
\begin{displaymath}
(2)(5) + (2)(5) + 10 - 7 - 20 \: = \: 3,
\end{displaymath}
as expected for a threefold.

We can describe (0,2) deformations of this theory as follows.
We take
\begin{eqnarray*}
E_{p_{ij}} &=& -\sigma \left( \tilde{N}^{\ell}_i p_{\ell j} - 
\tilde{N}^{\ell}_j p_{\ell i} \right), \\
E_{b^{ij}} &=& \sigma \left( N^i_k b^{k j} - N^j_k b^{k i} \right), \\
E_{\tilde{b}^{ij}} &= & \sigma \left( \tilde{N}^i_k \tilde{b}^{k j} - 
\tilde{N}^j_k \tilde{b}^{k i} \right), \\
E_{\varphi^a_i} &=& \sigma^a_b\varphi^b_i-\sigma N^j_i\varphi^a_j, \\
E_{\tilde{\varphi}^{a'}_i} &=& \tilde{\sigma}^{a'}_{b'}
\tilde{\varphi}^{b'}_i
- \sigma \tilde{N}^j_i\tilde{\varphi}^{a'}_j ,
\end{eqnarray*}
where $N^i_j$, $\tilde{N}^i_j$ are related by the 
constraint~(\ref{eq:first:constr}), namely
\begin{displaymath}
N^k_{[j_1} f^{i_1 i_2}_{j_2] k} \: = \:
\tilde{N}^{[i_1}_k f^{i_2] k}_{j_1 j_2},
\end{displaymath}
and for $J$'s:
\begin{eqnarray*}
J_{p_{ij}} &=& f^{ij}_{i^\prime j^\prime}b^{i^\prime j^\prime}-\tilde{b}^{ij}, \\
J_{b^{ij}} &=& p_{i_1i_2}f^{i_1i_2}_{ij}+\varphi^a_i\varphi^b_j J_{ab}, \\
J_{\tilde{b}^{ij}} &=& -p_{ij}+\tilde{\varphi}^a_i\tilde{\varphi}^b_j J_{ab}, \\
J_{\varphi^a_i} &=& 2 \varphi^{b}_{k}J_{a b}b^{i k}, \\
J_{\tilde{\varphi}^{a'}_i} &=& 2\tilde{\varphi}^{b'}_{k}J_{a' b'}
\tilde{b}^{i k} .
\end{eqnarray*}

It can be shown that for the deformations above, $E \cdot J=0$, using the
relation
\begin{equation}  \label{eq:symp}
\sigma^a_c J_{ab} \: = \: \sigma^a_b J_{ac},
\end{equation}
following from properties of the $Sp$ Lie algebra.

On the (2,2) locus,
\begin{displaymath}
N^i_j \: = \: \tilde{N}^i_j,
\end{displaymath}
just as in the original description.

\subsection{Single dual description}

In this section we dualize only one of the $SU(2)$ gauge factors of the
first model discussed to $Sp(N-3)$, giving gauge group
\begin{displaymath}
\frac{U(1)\times Sp(N-3)\times SU(2)}{\mathbb{Z}_2\times \mathbb{Z}_2}
\end{displaymath}
with 
\begin{itemize}
\item $N$ fields $\varphi^a_i$ in the $({\bf N-3},{\bf 1})_{-1}$
representation,
\item $(1/2) N (N-1)$ fields $b^{ij} = -b^{ji}$ in the 
$({\bf 1},{\bf 1})_2$ representation,
\item $N$ fields $\tilde{\phi}^j_{a'}$ in the
$({\bf 1},{\bf 2})_1$ representation,
\item $(1/2) N (N-1)$ fields $p_{ij} = - p_{ji}$ in the
$({\bf 1},{\bf 1})_{-2}$ representation,
\end{itemize}
with superpotential
\begin{displaymath}
W=\sum p_{ij}(f^{ij}(b)-\tilde{B}^{ij})+\varphi^a_i\varphi^b_jJ_{ab}b^{ij}
\end{displaymath}
where $\tilde{B}^{ij}=\epsilon^{a' b'}\tilde{\phi}^i_{a'} \tilde{\phi}^j_{b'}$ 
and $p_{ij}$ have charge $-2$, $b^{ij}$ have charge $2$, 
$\varphi^a_i$ have charge $-1$ and $\tilde{\phi}^i_{a'}$ 
have charge $1$ under the
$U(1)$ factor in the gauge group.      

In this duality frame, both the $r \gg 0$ and $r \ll 0$ phases have
geometry determined in part by nonperturbative effects as in \cite{Hori:2006dk},
so as mentioned in the introduction, at this
time we cannot
provide a simple monad description of either.

In $(0,2)$ language, $E$ deformations are
\begin{eqnarray*}
E_{p_{ij}} & =& -\sigma \left( \tilde{N}^k_i p_{kj} - \tilde{N}^k_j p_{ki}
\right), \\
 E_{b^{ij}} &= & \sigma \left( N^i_k b^{kj} - N^j_k b^{ki} \right), \\
E_{\varphi^a_i} &=& \sigma^a_b\varphi^b_i-\sigma N^j_i\varphi^a_j, \\
 E_{\tilde{\phi}^i_{a'}} &=& \tilde{\sigma}^{b'}_{a'} \tilde{\phi}^i_{b'}
+ \sigma\tilde{N}^i_j\tilde{\phi}^j_{a'} ,
\end{eqnarray*}
where $N^i_j$, $\tilde{N}^i_j$ are related by the same
constraint~(\ref{eq:first:constr}) as in the last two duality frames, namely
\begin{displaymath}
N^k_{[j_1} f^{i_1 i_2}_{j_2] k} \: = \:
\tilde{N}^{[i_1}_k f^{i_2] k}_{j_1 j_2},
\end{displaymath}
and $J$ deformations are
\begin{eqnarray*}
J_{p_{ij}} &=& f^{ij}(b)-\tilde{B}^{ij}, \\
 J_{b^{ij}} &=& p_{mn}f^{mn}_{ij} + \varphi^a_i \varphi^b_j J_{ab}, \\
J_{\varphi^a_i} &=& 2\varphi^{b}_{j}J_{a b}b^{i j}  , \\
J_{\tilde{\phi}^i_{a'}} &=& 
-2p_{i j}\epsilon^{a' b'}\tilde{\phi}^j_{b'}.
\end{eqnarray*}
It is straightforward to show that $E J = 0$,
using the symplectic property~(\ref{eq:symp}) of $\sigma^a_b$ and the
tracelessness of $\tilde{\sigma}^{a'}_{b'}$.

Just as in the last two duality frames, on the (2,2) locus,
\begin{displaymath}
N^i_j \: = \: \delta^i_j \: = \:
\tilde{N}^i_j.
\end{displaymath}

\subsection{Comparison of deformations}

It is tempting to identify the $N$ in any one duality frame with the
$N$ in any other, and similarly the $\tilde{N}$ in any one duality frame
with the $\tilde{N}$ in any other.  As a cautionary note, however, we observe
that this is potentially too simplistic.  For example, 
(0,2) deformations of a Grassmannian $G(k,n)$ are parametrized by
\cite{Guo:2015caf,Guo:2016suk}
an $n \times n$ matrix $B^i_j$, and one might guess that the
transpose would give the corresponding (0,2) deformations of
the dual Grassmannian $G(n-k,n)$.
However,
as observed in
\cite{Guo:2015caf}, merely taking the transpose of $B^i_j$ does not
generate the quantum sheaf cohomology ring of the deformed dual Grassmannian,
hence the correct parameter match is more complicated than merely
taking the transpose.  In the present case, it is possible that the
correct parameter match is more complicated than merely identifying
all instances of $N^i_j$ and $\tilde{N}^i_j$ in different duality frames.

\section{Examples of anomaly-free bundles on $G(2,5) \cap G(2,5)$}
\label{sect:exs-anom-free}

In table~\ref{table:anom-free}, we summarize several anomaly-free bundles
built as kernels, in the form
\begin{displaymath}
0 \: \longrightarrow \: {\cal E} \: \longrightarrow \: {\cal O}(A) 
\: \longrightarrow
\: {\cal O}(B) \: \longrightarrow \: 0,
\end{displaymath}
where $A$ and $B$ are representations of the gauge group.

\begin{table}
\begin{center}
\begin{tabular}{ccc}
rank & $A$ & $B$ \\ \hline
6 & $({\bf 2},{\bf 1})_{-1} \oplus ({\bf 1},{\bf 2})_{-1} \oplus ({\bf 1},{\bf 1})_2 \oplus ({\bf 1},{\bf 1})_6^{\oplus 2}$ &
$({\bf 1},{\bf 1})_{10}$ \\
7 & $({\bf 2},{\bf 1})_{-1} \oplus ({\bf 1},{\bf 2})_{-1} \oplus
({\bf 1},{\bf 1})_2^{\oplus 2} \oplus ({\bf 1},{\bf 1})_4^{\oplus 2}$
& $({\bf 1},{\bf 1})_8$ \\
7 & $({\bf 2},{\bf 1})_{-1} \oplus ({\bf 1},{\bf 2})_{-3} \oplus 
({\bf 1},{\bf 1})_4^{\oplus 2} \oplus ({\bf 1},{\bf 1})_6^{\oplus 2}$
& $({\bf 1},{\bf 1})_{12}$ \\
7 & $({\bf 2},{\bf 1})_{-3} \oplus ({\bf 1},{\bf 2})_7 \oplus 
({\bf 1},{\bf 1})_4 \oplus ({\bf 1},{\bf 1})_6 \oplus
({\bf 1},{\bf 1})_{-2}^{\oplus 2}$ 
& $({\bf 1},{\bf 1})_{14}$ 
\end{tabular}
\end{center}
\caption{\label{table:anom-free} A few anomaly-free bundles on
$G(2,5) \cap G(2,5)$.}
\end{table}

Each of these bundles is represented by a (0,2) GLSM with gauge group
\begin{displaymath}
\frac{
U(1) \times SU(2) \times SU(2)
}{
{\mathbb Z}_2 \times {\mathbb Z}_2
}
\end{displaymath}
with matter
\begin{itemize}
\item $5$ chiral superfields $\phi^i_a$ in the $({\bf 2},{\bf 1})_1$ representation,
\item $5$ chiral superfields $\tilde{\phi}^i_{a'}$ in the $({\bf 1},{\bf 2})_1$ 
representation,
\item $10$ Fermi superfields $\Lambda_{ij} = - \Lambda_{ji}$ in the
$({\bf 1},{\bf 1})_{-2}$ representation,
\item Fermi superfields $\Lambda^{\alpha}$ in representation $A$,
\item chiral superfields $P_{\beta}$ in the dual of representation $B$,
\end{itemize}
with (0,2) superpotential
\begin{displaymath}
\Lambda_{ij} \left( f^{ij}(B) - \tilde{B}^{ij} \right)
\: + \: P_{\beta} J^{\beta}_{\alpha} \Lambda^{\alpha}.
\end{displaymath}
It is straighforward to check that each representation in
table~\ref{table:anom-free} is invariant under ${\mathbb Z}_2 \times
{\mathbb Z}_2$, where each ${\mathbb Z}_2$ factor relates the center of
one $SU(2)$ to a subgroup of $U(1)$.

In addition, there is also an anomaly-free bundle defined similarly by
the data
\begin{center}
\begin{tabular}{ccc}
rank & $A$ & $B$ \\ \hline
4 & $({\bf 2},{\bf 2})_4 \oplus ({\bf 1},{\bf 1})_2^{\oplus 3} \oplus
({\bf 1},{\bf 1})_{-2}$
& $({\bf 2},{\bf 1})_5 \oplus ({\bf 1},{\bf 2})_5$ 
\end{tabular}
\end{center}
in the GLSM with gauge group
\begin{displaymath}
U(1) \times SU(2) \times SU(2)
\end{displaymath}
and the same matter and superpotential as above.
(Here, representation $A$ is not invariant under the
${\mathbb Z}_2 \times {\mathbb Z}_2$.)

In each case,
Green-Schwarz anomaly cancellation requires, schematically,
\begin{displaymath}
\sum_{ R_{\rm left} } {\rm tr}\left( T^a T^b \right) \: = \:
\sum_{ R_{\rm right} } {\rm tr}\left( T^a T^b \right).
\end{displaymath}
Anomaly cancellation for the $U(1)$ factor can be understood in the
standard form, as a sum of squares of charges.  For the nonabelian
factors, anomaly cancellation in each factor can be written more explicitly as
\cite{Jia:2014ffa}[section 3.1]
\begin{displaymath}
\sum_{R_{\rm left}} \left( {\rm dim}\, R_{\rm left}\right)
{\rm Cas}_2\left( R_{\rm left} \right) \: + \:
\left( {\rm dim}\, {\rm adj}\right) {\rm Cas}_2\left( {\rm adj} \right)
\: = \:
\sum_{ R_{\rm right} } \left( {\rm dim}\, R_{\rm right} \right)
{\rm Cas}_2\left( R_{\rm right} \right),
\end{displaymath}
where we have explicitly incorporated the left-moving gauginos into the
expression above.
As $SU(2)$ generators are traceless, there are $SU(2)-SU(2)$ and
$SU(2)'-SU(2)'$ anomalies, but no $U(1)-SU(2)$ or $SU(2)-SU(2)'$
anomalies to check.
In checking such anomalies, it is handy to note that for an $n$-dimensional
representation of $SU(2)$, Cas$_2$ is given by 
\cite{iachello}[equ'n (7.27)]
\begin{displaymath}
\frac{1}{2}\left( n^2 - 1 \right)
\end{displaymath}
(using the fact that $\lambda_1=n-1$ in that reference's conventions).

\section{Conclusions}

In this paper we have examined (0,2) deformations and properties of
some exotic GLSMs.  We have seen that in GLSMs realizing branched
double covers nonperturbatively, the tangent bundle is realized as the
cohomology of three-term sequence involving both bundles and torsion sheaves,
a novel effect in (0,2) theories.  We have also seen several examples in 
which quasi-isomorphisms arise physically, relating monads in theories
at different points along RG flow, their first occurrence outside of
applications to B-branes and derived categories.  Finally, we have
examined (0,2) deformations in the various duality phases of 
the intersection $G(2,N) \cap G(2,N)$, and have listed a few examples
of anomaly-free (0,2) theories on $G(2,5) \cap G(2,5)$.

\section{Acknowledgements}

We would like to thank A.~C\u{a}ld\u{a}raru, J.~Guo,
J.~Knapp, Z.~Lu, and T.~Pantev for useful discussions.
E.S. was partially supported by NSF grant PHY-1720321.

\end{document}